\documentclass[fleqn,usenatbib]{mnras}

\usepackage{newtxtext,newtxmath}
\usepackage{xcolor}

\usepackage[T1]{fontenc}

\DeclareRobustCommand{\VAN}[3]{#2}
\let\VANthebibliography\thebibliography
\def\thebibliography{\DeclareRobustCommand{\VAN}[3]{##3}\VANthebibliography}

\usepackage{graphicx}	
\usepackage{amsmath}	

\title[Synchrotron Signatures of CR Transport]{Synchrotron Signatures of Cosmic Ray Transport Physics in Galaxies}

\author[S. Ponnada et al.]{
\vspace{0.1cm}
\parbox[t]{\textwidth}{Sam B. Ponnada,$^{1}$\thanks{E-mail: sponnada@astro.caltech.edu}
Iryna S. Butsky,$^{1,2}$
Raphael Skalidis,$^{1}$
Philip F. Hopkins,$^{1}$
Georgia V. Panopoulou,$^{3}$
Cameron Hummels,$^{1}$
Du\v{s}an Kere\v{s},$^{5}$
Eliot Quataert,$^{4}$
Claude-Andr\'e Faucher-Gigu\`ere,$^{6}$
Kung-Yi Su$^{7}$}
\\
$^{1}$California Institute of Technology, TAPIR, Mailcode 350-17, Pasadena, CA 91125, USA\\
$^{2}$Kavli Institute for Particle Astrophysics \& Cosmology (KIPAC), Stanford University, Stanford, CA
94305, USA\\
$^{3}$Department of Space, Earth and Environment, Chalmers University of Technology, 412 93, G\"oteborg, Sweden\\
$^{4}$ Department of Astrophysical Sciences, Princeton University, Princeton, NJ 08544, USA \\
$^{5}$ 
Department of Physics, Center for Astrophysics and Space Sciences, University of California San Diego
, 9500 Gilman Drive, La Jolla, CA 92093, USA\\
$^{6}$Department of Physics and Astronomy and CIERA, Northwestern University, 2145 Sheridan Road, Evanston, IL 60208, USA\\
$^{7}$Black Hole Initiative, Harvard University, 20 Garden Street, Cambridge, MA 02138, USA\\
}

\date{Accepted 2024 February 27. Received 2024 February 21; in original form 2023 September 28}

\pubyear{2024}

\begin{document}
\label{firstpage}
\pagerange{\pageref{firstpage}--\pageref{lastpage}}
\maketitle

\begin{abstract}
Cosmic rays (CRs) may drive outflows and alter the phase structure of the circumgalactic medium, with potentially important implications on galaxy formation. However, these effects ultimately depend on the dominant mode of transport of CRs within and around galaxies, which remains highly uncertain. To explore potential observable constraints on CR transport, we investigate a set of cosmological FIRE-2 CR-MHD simulations of L$_{\ast}$ galaxies which evolve CRs with transport models motivated by self-confinement (SC) and extrinsic turbulence (ET) paradigms. To first order, the synchrotron properties diverge between SC and ET models due to a CR physics driven hysteresis. SC models show a higher tendency to undergo `ejective' feedback events due to a runaway buildup of CR pressure in dense gas due to the behavior of SC transport scalings at extremal CR energy densities. The corresponding CR wind-driven hysteresis results in brighter, smoother, and more extended synchrotron emission in SC runs relative to ET and constant diffusion runs. The differences in synchrotron arise from different morphology, ISM gas and \textbf{B} properties, potentially ruling out SC as the dominant mode of CR transport in typical star-forming L$_{\ast}$ galaxies, and indicating the prospect for non-thermal radio continuum observations to constrain CR transport physics.
\end{abstract}

\begin{keywords}
ISM: cosmic rays -- ISM: magnetic fields -- galaxies: formation -- methods: numerical
\end{keywords}



\section{Introduction}
Relativistic charged particles, or cosmic rays (CRs), are ubiquitous in the Universe. Injected and accelerated at supernovae (SNe), stellar winds, and associated shocks fronts, CRs are known to be a considerable component of the Milky Way (MW) interstellar medium (ISM) \citep{Boulares1990,Bell1978} and are observed in other L$_{\ast}$ galaxies via their $\gamma$-ray and non-thermal synchrotron radiation \citep{lacki_gev_2011,tang_discovery_2014}. 

In the past decade, the importance of CRs as a source of feedback in galaxies has come to be appreciated \citep[for recent reviews, see][]{owen_cosmic_2023,ruszkowski_cosmic_2023}. A host of theoretical studies employing varied numerical and physical prescriptions have established that CRs can play an important role in driving and altering the structure of winds \citep{booth_simulations_2013,salem_cosmic_2014,girichidis_launching_2016,simpson_role_2016,pakmor_galactic_2016,bustard_cosmic-ray-driven_2020,huang_launching_2022,huang_cosmic-ray-driven_2022,quataert_physics_2022,armillotta_cosmic-ray_2022,thomas_cosmic-ray-driven_2023,modak_cosmic-ray_2023} and providing a potentially key source of non-thermal pressure support in the circum-galactic medium (CGM) \citep{Butsky2018,chan_cosmic_2019, buck_effects_2020, Hopkins2020,farcy_radiation-magnetohydrodynamics_2022}. 

These effects can manifestly change the star formation histories of L$_{\ast}$ galaxies by preventing cool gas from precipitating onto the disk, altering the dynamics of gas in the tenuous inner CGM \citep{Butsky2022} or `disk-halo interface` \citep{Chan2021} with potential implications on the amplification of magnetic fields \citep{Ponnada2022} as well as the phase structure and ionization state of halo gas \citep{salem_role_2016,Ji2020,butsky_impact_2020,tsung_impact_2023}.

However, a major caveat remains that all of the aforementioned effects depend sensitively on the dominant mode of transport of CRs through the ISM and into the CGM, which is highly uncertain with elusive observational constraints \citep{hopkins_effects_2021}. An understanding of CR transport is thus crucial to contextualize the importance of CRs for galaxy formation and evolution, as CR effects in the ISM and CGM are heavily dependent on the macroscopic transport speed, often parameterized through the diffusion coefficient $\kappa$ (more specifically, $\kappa_{\rm \|}$), or streaming speed v$_{\rm st}$. 

 The transport of CRs on $\sim$kpc-Mpc galactic scales is fundamentally tied to the scattering of CRs on orders-of-magnitude smaller gyro-resonant scales ($\sim$ 0.1 AU for $\sim$GeV CRs). Thus, there has been increasing theoretical interest in understanding the macro-physical transport properties of CRs motivated by models of plasma-scale CR transport \citep{jokipii_cosmic-ray_1966,skilling_cosmic_1975} and how their predicted observables compare to observations \citep{hopkins_testing_2021,hopkins_standard_2022,kempski_reconciling_2022,butsky_constraining_2023}. 

Despite some constraining power of existing observations, there is a dire need for further observational comparison to narrow the broad theoretical parameter space, which radio-continuum synchrotron observations may provide. In this Letter, we forward-model synchrotron emission from cosmological, zoom-in simulations of galaxy formation including CRs with different physically-motivated CR transport models from the Feedback in Realistic Environments (FIRE) suite\footnote{\url{https://fire.northwestern.edu/}} \citep{Hopkins2018,hopkins_testing_2021} and explore the physical basis for corresponding observable differences which emerge owing to CR physics. In Section \ref{sec:methods}, we briefly describe the simulations and our methods. Then, we present our results for models with varied CR transport physics in Section \ref{sec:cr_variants}. Lastly, we discuss our conclusions in Section \ref{sec:discussion}.

\section{Simulations and methods}\label{sec:methods}
In this study, we utilize a subset of the simulations presented in \citep{hopkins_effects_2021,hopkins_testing_2021} which evolve a `single-bin' of 1-10 GeV CRs and utilize FIRE-2 \citep{Hopkins2018} physics. We summarize the most pertinent aspects here, but refer the reader to the aforementioned papers for a more in-depth discussion of numerical details.

The simulations are all fully cosmological, magnetohydrodynamic \citep{hopkins_accurate_2016,hopkins_constrained-gradient_2016} simulations of galaxy formation which include baryons and dark matter, fully anisotropic Spitzer-Braginskii conduction and viscosity \citep{hopkins_anisotropic_2017} at a Lagrangian mass resolution of 56000 M$_{\rm \odot}$. Prescriptions for explicit stellar feedback and gas cooling (for T $\sim$ 10-10$^{10}$ K) follow \citep{Hopkins2018}; stars form in dense (n $>$ 1000 cm$^{-3}$), self-shielded, Jeans unstable gas with multi-band radiation, mass-loss, and explosive feedback from Types Ia and II SNe (evolved self-consistently following stellar evolution models) coupled to gas. 

Cosmic rays are injected from SNe and OB/WR stellar winds with an energy efficiency of  $\epsilon_{\rm CR} = $ 0.1 of the inital ejecta kinetic energy. In these `single-bin' simulations, we solely evolve the $\sim$1-10 GeV CR energy density (e$_{\rm CR}$), or equivalently a constant spectral distribution, as a relativistic fluid with $\gamma_{\rm CR} =$ 4/3. The CR dynamics are coupled to the gas and evolve self-consistently, with transport coupled to magnetic field lines according to the CR transport equations and loss terms (collisional, streaming) computed in-code \citep[again, see details in][]{hopkins_testing_2021}.

These simulations invoke scalings for the CR scattering rate, $\nu$, with various plasma properties motivated by micro-physical scenarios. One such model class includes "extrinsic turbulence" (ET) scenarios \citep{jokipii_cosmic-ray_1966}, where CRs are scattered off of gyro-resonant fluctuations in \textbf{B} on scales of order the CR gyro-radius that arise from a turbulent cascade down to those (small) scales. Model variants in this general class vary widely (as shown in \citealt{hopkins_testing_2021}) according to uncertainties in the shape of the turbulent cascade at small scales, which turbulent modes are of primary importance for scattering on these scales, the importance of certain damping terms, and geometric considerations of the (an)isotropy of said turbulent modes. But broadly speaking, the assumption for our purposes is that the scattering rate $\nu$ varies with the local Alfv\'en scale ($\ell_{\rm A}$) and Alf\'ven Mach number ($\mathcal{M}_{\rm A}$) of turbulence on \textit{resolved} simulation scales as $\nu \propto \mathcal{M}_{\rm A}^{2}/\ell_{\rm A}$. The normalization of $\nu$ for these models at $\sim$1 GeV is fitted by \citet{hopkins_testing_2021}
to the Voyager, AMS-02, and Fermi data.

The second primary class of models are "self-confinement" scenarios \citep{skilling_cosmic_1975}, in which CRs excite Alfv\'en waves as they stream down their pressure gradients, which dominates the generation of gyro-resonant fluctuations in \textbf{B} which subsequently scatter CRs. The CR scattering is determined by the balance of the growth and damping of these gyro-resonant Alfv\'en waves and so model variants within this class are sensitive to the choice of Alfv\'en speed, assumptions regarding the wave damping and growth terms, and uncertainties in the turbulent dissipation timescales. The key scaling here for ultra-relativistic CRs is $\nu \propto (\frac{e_{\rm CR}}{e_{\rm B}})(\frac{\mathrm{v}_{\rm A} \mathrm{c}}{\ell_{\rm CR}r_{\rm L} \Gamma})$ in terms of the magnetic and CR energy densities e$_{\rm B}$, e$_{\rm CR}$; Alfv\'en speed v$_{\rm A}$; gradient scale length $\ell_{\rm CR}$; gyro radius r$_{\rm L}$; and plasma damping terms $\Gamma$. These are again re-normalized in \citet{hopkins_testing_2021} to fit the aforementioned $\sim$1-10 GeV observations. 

The subset of model variants from \citet{hopkins_testing_2021} explored here were shown to reasonably reproduce observables of $\gamma$-ray emission, effective isotropic diffusivities, and cosmic ray energy densities at the "Solar circle", though we will also describe results for simulations which were not consistent with the above constraints to illustrate qualitative differences tying the physics of the model class to the synchrotron properties. 

We also compare these model variants to a FIRE-2 simulation that uses a spatially and temporally constant scattering rate (hereafter called the 'constant diffusivity' or CD run) presented in \citet{Hopkins2020}, and whose magnetic field properties were detailed extensively in \citet{Ponnada2022}. This run's constant parallel diffusivity is $\kappa_{\|}$ = 3 $\times$ 10$^{29}$ cm$^{2}$/s, which was chosen to be consistent with the aforementioned constraints \citet{chan_cosmic_2019}.

To generate our synchrotron predictions, we follow the procedure outlined in \citet{ponnada_synchrotron_2024}, with the caveat that as these are `single-bin' simulations, we assume a constant CR electron (CRe) spectral shape of \citet{bisschoff_new_2019} and scale the spectrum by the ratio of each gas cell's self-consistently evolved e$_{\rm CR}$ to the local ISM value. This is akin to assuming a constant proton-to-electron ratio as well as a constant spectral shape. Since \citet{bisschoff_new_2019} provides an empirical spectrum, we are assuming that these models have been tuned to give the right spectral slope according to constraints at Milky Way Solar Circle, though see \citet{kempski_reconciling_2022} and \citet{hopkins_standard_2022} for why this may not be physically possible in practice. 

Subsequently, the following analysis cannot capture the effects of potential variation in spectral shape and proton-to-electron ratios owing to varying CRe loss terms in gas of different phases and ionization states, nor variation owing to the varied CR transport models and their coupling to gas properties. However, this provides a first look at how the emission properties differ to first-order owing to dynamical differences and corresponding effects on phase structure and gas properties owing to CR transport effects, notwithstanding the caveats mentioned above.

\section{Synchrotron Emission and The Physics of Cosmic Ray Transport}\label{sec:cr_variants}
\begin{figure*}
    \centering
    \includegraphics[width=1.0\textwidth]{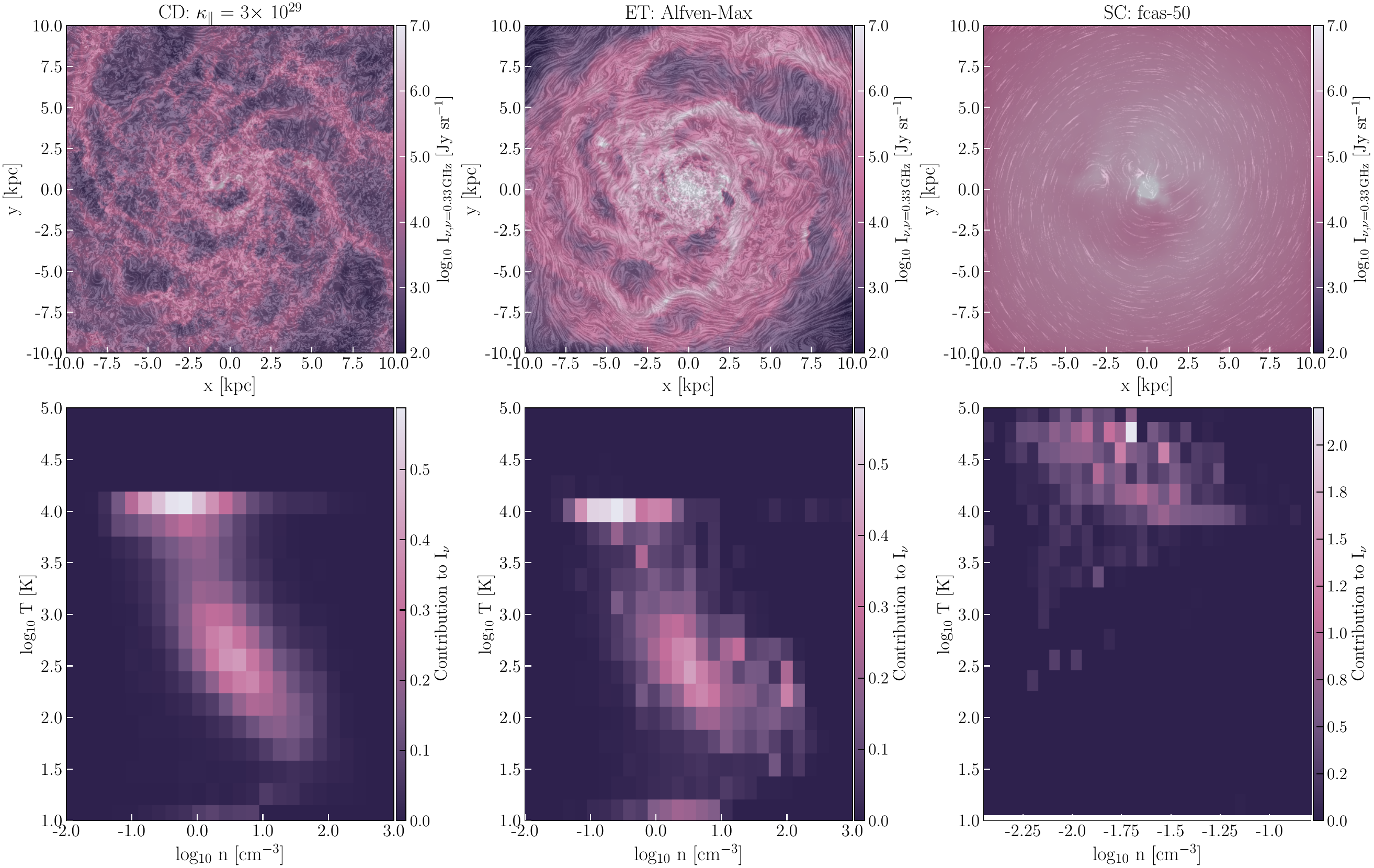}
    \caption{\textit{Visualizations of the synchrotron emission at 0.33 GHz and intensity-weighted phase diagrams for FIRE-2 simulations of \texttt{m12i} with varied CR transport physics at $z$ = 0.} \textbf{Row 1:} Specific intensity maps with superimposed lines showing the orientation of the mass-averaged components of the magnetic field. A model variant with spatially and temporally constant $\kappa_{\rm \|}$ = 3 $\times$ 10$^{29}$ cm s$^{-1}$ is shown on the left, a variant within the ET class of models (`Alfv\'en-Max') is shown in the middle, and a SC model ('fcas-50') on the right. CD and ET models generally exhibit more turbulent structure in the magnetic fields, weaker emission, and more variation in brightness contrast to highly ordered \textbf{B} and brighter and smoother emission in the SC models. \textbf{Row 2:} Intensity-weighted histograms for 2 $<$ R/kpc $<$ 10 and $|$z$|$ $<$ 3 kpc for the CD, ET and SC runs above. We exclude the central 2 kpc in order to characterize the extended emission properties rather than the bright central cores. In SC models, the synchrotron emission primarily arises from the WIM/WNM compared to the CNM/WNM dominated scenario in the CD and ET runs.}
    \label{fig:cr_variant_viz}
\end{figure*}

We examine the synchrotron emission and magnetic field structure from two representative model variants in the ET and SC model classes in Figure \ref{fig:cr_variant_viz} and characterize key differences in the properties of the gas giving rise to the emission. 

There appears to be a dichotomy, on average, in the physical morphologies of the galaxies in the two model classes. ET runs exhibit more typical spiral structure and SC runs have a more central bulge-dominated, lenticular-like appearance. The SC runs tend to show brighter, smoother, and more extended emission and have more ordered magnetic field structure relative to the ET runs; ET runs look qualitatively similar to the constant diffusivity run, with brighter emission coincident with the spiral arms and neutral gas structures in the galactic center. The physical differences underpinning the visual differences between the ET and SC runs become clear in the intensity weighted histograms (Figure \ref{fig:cr_variant_viz}, bottom panels).  Figure \ref{fig:cr_variant_viz} shows that the extended emission in the ET runs is primarily arising from the denser cool and warm neutral gas while the SC runs have emission mostly arising from warmer and more diffuse gas.

\begin{figure}
    \centering
    \includegraphics[width=0.5\textwidth]{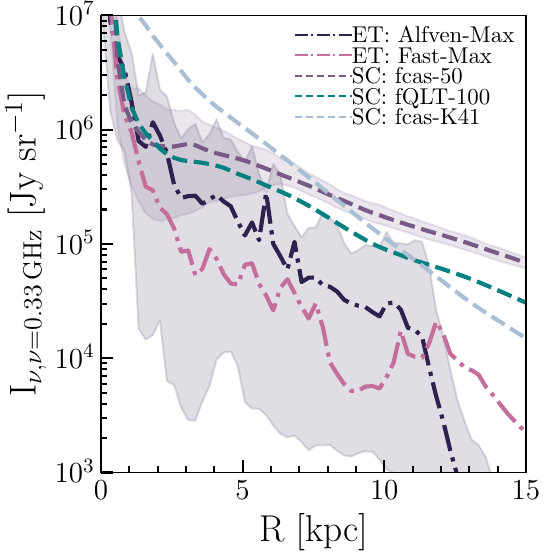}
    \caption{\textit{Azimuthally averaged, face-on radial profiles of synchrotron specific intensity for FIRE-2 simulations of \texttt{m12i} with varied CR transport physics at $z$ = 0}. Lines show simulations with ET (dot-dashed) and SC (dashed) model variants of CR transport. Shaded regions show the 5-95 percent range at a given radial bin. Our predictions show significant differences in the shape and normalization of the synchrotron emission profiles, with pathologically different behaviors exhibited between model classes. SC models tend to show brighter, smoother, and more extended profiles in comparison to ET and CD models. The difference in the profiles arises qualitative differences in the phase structure, magnetic field properties, and gas distribution modulated by a CR-physics driven hysteresis.} 
    \label{fig:cr_variant_profiles}
\end{figure}

In Figure \ref{fig:cr_variant_profiles}, we examine these differences more quantitatively with radial profiles of the forward-modeled synchrotron emission for CR physics model variants simulated in \citet{hopkins_testing_2021} that met their reasonable observational $\gamma$-ray and e$_{\rm CR}$ constraints. We see significant variation in the profiles depending on CR transport physics. We see a separation between the ET and SC model variants: SC runs typically exhibit brighter emission averaged at a given radius by a factor of $\sim$3-10 relative to ET runs, despite brighter clumped peaks in the spiral arms of ET runs. The SC runs also exhibit smoother emission that falls off more gradually with radius relative to ET and constant diffusivity runs. We stress that the correlation is \textit{not} one-to-one; we can see many earlier (higher-redshift) snapshots where the SC models look more like ET. And some simulations with very low constant diffusivity (2 dex lower than observationally allowed) look similar to the SC runs. We discuss this below.

While the radial profiles for the SC runs appear to be qualitatively more similar to a couple of the known observational profiles in that they exhibit a shallower falloff with radius \citep{basu_magnetic_2013,beck_magnetic_2015}, the apparent morphological features of the galaxies look markedly different. We defer a comprehensive observational comparison to future work using spectrally-resolved cosmological runs. The variation in the synchrotron profiles between classes of CR transport models indicate the potential for the comparison of larger samples of spatially resolved synchrotron images to model predictions to constrain deeply uncertain CR transport physics.

The shape, normalization, and scatter in the profiles is a function of the phase of the ISM dominating the galaxy. The smoothness of the SC profiles is induced by the emission arising mostly from the warm neutral/warm ionized media (WNM/WIM), while on the other hand, the synchrotron intensity profiles of the ET and CD runs are dominated by emission coming from the WNM and denser cold neutral medium (CNM). This key physical difference appears to be driven by differences in the CR transport physics between the SC and ET models, as we will describe in the next section.

\subsection{A Cosmic Ray Physics Driven Hysteresis}

The striking differences between the observables and properties of the CD, ET and SC models boil down to some crucial differences in the physics of CR transport. One of the main features of SC models is the (general) scaling of the scattering rate (see Section \ref{sec:methods}) as $\nu \propto e_{\rm CR}$, i.e., the effective/emergent diffusion coefficient is inversely proportional to e$_{CR}$ ($\kappa_{\|} \propto e_{\rm CR}^{-1}$ in SC model variants, which is the defining characteristic of these types of models; for exact scalings of the models variants considered, see \citealt{hopkins_testing_2021}). This scaling is true when the linear damping term dominates the gyro-resonant Alf\'ven waves, and the CR flux is in approximate local steady state. This inverse scaling of the diffusion coefficient with the CR energy density can lead to scenarios in which regions of high e$_{\rm CR}$ are prone to more efficient trapping of CRs. This trapping of CRs then leads to the limit of increasing e$_{\rm CR}$, therefore increasing $\nu$ and so on until $\nu \rightarrow \infty$, and the CRs are trapped to move strictly with Alfv\'en wave packets in the gas. This means a large CR pressure has built up and been "trapped" in the dense ISM gas. This build-up of CR pressure eventually blows apart ISM gas, and thus the galaxy is largely filled with warm/hot and diffuse phases, with dense, magnetized, CR-laden gas spread via these outflows into a much larger, smoother distribution. In contrast, regions of high e$_{\rm CR}$ in ET runs would rapidly diffuse/escape, and due to high e$_{\rm CR}$ compressive modes can be effectively damped, even further "de-confining" CRs locally.

This difference in the behavior of CRs especially at high e$_{\rm CR}$ seems to underpin a CR physics driven hysteresis between the SC model variants and the rest. In SC runs, at $z = 0$ we typically see a warmer and more diffuse phase structure, lower gas surface densities outside R $\sim$4 kpc, stronger and more ordered \textbf{B} at a given $n_{\rm gas}$ and at a given radius, and a steeper e$_{\rm CR}$ - $n_{\rm gas}$ relation. These differences primarily appear to arise after a non-linear feedback event owing to the SC-runaway in which CRs expel most of the cool and neutral gas outside of R $\sim$4 kpc. At the earlier snapshots this has not yet occurred; it is of course possible that no runaway occurs, but based on the analysis of the 3 SC-motivated runs that meet constraints here, as well as 6 of the SC-motivated runs that failed to meet constraints in \citep{hopkins_testing_2021}, we conclude that it happens eventually 
than not, as we do not see \textit{any} SC-motivated runs that do not suffer from this issue. 

\begin{figure}
    \centering
    \includegraphics[width=0.5\textwidth]{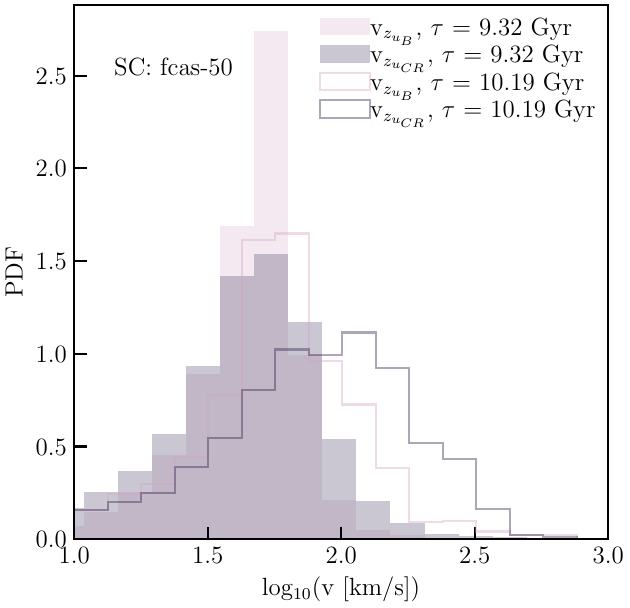}
    \caption{\textit{PDFs of the gas velocity log$_{10}$($|$v$_{\rm z}$$|$) weighted by u$_{\rm B}$ (pink) and e$_{\rm CR}$ (black)} at two snapshots 820 Myr apart (filled and unfilled) for R $<$ 14 kpc at heights from the mid-plane of 0.5-3 kpc for a SC run (`fcas-50'). Runs with SC model variants for CR transport appear to be more likely to undergo extreme feedback scenarios in which a build-up of e$_{\rm CR}$ runs away until expelling highly magnetized and e$_{\rm CR}$-loaded winds from the galaxy. These winds carry away cool, neutral gas and transform the phase structure and corresponding observable properties of the synchrotron emission.}
    \label{fig:v_hists}
\end{figure}

To see this in more detail, in Figure \ref{fig:v_hists}, we show PDFs of the vertical component of velocity ($|v_{\rm z}|$) weighted by u$_{\rm B}$ and e$_{\rm CR}$ for two snapshots $\sim$820 Myr apart of the SC run `fcas-50' at displacements of 0.5-3 kpc from the disk mid-plane. The later snapshot has clear signatures of a feedback event, with the e$_{\rm CR}$-weighted velocity PDF shifting to having many gas cells with $|v_{\rm z}|$ $>$ 100 km/s, and the magnetic energy density-weighted PDF shifting similarly, though with lower magnitude. The presence of these e$_{\rm CR}$-loaded winds corresponds directly with a transition in these SC runs from morphological spirals with relatively similar gas distributions, ISM phase structure, and magnetic field properties to the ET and CD runs.

While we show only the velocity PDFs for `fcas-50', this general picture of e$_{\rm CR}$-loaded winds, which drive substantial changes in the galaxy properties and synchrotron observables appears to emerge for the other SC models explored in this paper as well. As further confirmation of this process, we note that we see a similar effect of CR and u$_{\rm B}$-loaded winds from "trapped" CRs in runs not shown here but run in \citet{hopkins_testing_2021} where they adopted a constant but extremely large scattering rate (very low diffusivity, factors $>100$ lower than the observationally-allowed values). As noted by those authors, those particular runs were strongly ruled out by CR spectra, primary-to-secondary ratios, and $\gamma$-ray emission in the Galaxy, hence our not comparing them further here. But, by definition, they produce efficient CR trapping, so it should not be surprising that they can produce a similar "blowout" event to the SC runs here. This demonstrates a new prediction for variations of CR transport models in the SC regime: if CR transport at 1-10 GeV is dominated by modulation from self-excited, gyro-resonant Alfv\'en waves, galaxies may be more conducive to `ejective feedback' scenarios through CR-driven winds.

\section{Discussion and Conclusions}\label{sec:discussion}
In this work, we explore the effects of different physically-motivated models for the CR scattering rate $\nu$ which allow it to vary dynamically as function of local plasma properties, heuristically motivated by self-confinement (SC) and extrinsic turbulence (ET) models, in "single-bin" simulations (not evolving the full CR spectrum) calibrated to give reasonable mean $\langle \nu \rangle$ at $\sim$GeV energies in \citet{hopkins_testing_2021}.

Simulated galaxies with SC models of CR transport tend to have brighter, more spatially extended and smoother synchrotron emission than ET and CD models. The brighter emission in the SC models corresponds with a relatively featureless, warm-hot phase dominated ISM, elevated \textbf{B}-n$_{\rm gas}$ relation, and a more ordered and mean-field dominated \textbf{B}.This apparent hysteresis seems to be CR physics driven, as SC runs have the potential for a runaway at high e$_{\rm CR}$ which leads to CR energy concentrating until cold and dense gas is blown out via e$_{\rm CR}$ and u$_{\rm B}$ loaded winds, resulting in the stark morphological and physical differences between SC and ET/CD runs. 

Already, the sheer lack of detailed cold, neutral phase structure diverges from typical $\sim$ L$_{\ast}$ spiral galaxies, which may indicate that SC is not the dominant mode of CR transport in these types of galaxies, though it may operate more so within galaxies with a lenticular-like morphology with a more featureless gas/dust distribution. Despite this, the radial intensity profiles of the SC models are characteristically less steep than those of CD/ET models, and more similar in shape to the small sample of observed radial profiles compared to in \citet{ponnada_synchrotron_2024}. This may also indicate that something is missing from ET models, but we have not found a way to hybridize this model class with SC scalings in a way that does not suffer the `SC runaway' effects. It is easier in principle to reconcile the relative steepness of ET synchrotron emission profiles with physics not directly related to the CR transport scalings through slightly higher gas surface densities or magnetic field strengths. However, this work indicates the potential for differences between CR transport models to be probed in a spatially resolved manner with larger samples with future radio instruments like the DSA-2000 \citep{DSA2000}, ngVLA \citep{Murphy2018}, and Square Kilometer Array \citep{SKA} and with already existing and future HI 21 cm surveys \citep{Walter2008}. 

 We emphasize also that the differences seen in the model variations here are highly nonlinear, and do not indicate that SC models of CR transport will \textit{always} exhibit these differences relative to ET/CD models. Rather, the predictions made here are for SC transport models which have undergone the `SC runaway,' and simulations which have not undergone this nonlinear process (like higher redshift snapshots or those not run with the SC transport scalings fully cosmologically) do not exhibit the same characteristic synchrotron properties. And we stress that, as shown in more detail in \citet{hopkins_testing_2021,hopkins_standard_2022}, qualitative and order-of-magnitude uncertainties remain in first-principles models for the CR scattering rate $\nu$ and indeed \textit{no} first-principles model has been demonstrated to predict the correct CR spectra and primary-to-secondary ratios at $\sim$MeV-TeV energies \citep{hopkins_standard_2022}.
 
 And although the differences explored here appear to be driven by the CR physics, there are several other interrelated factors that may be important. Notably, the non-linear interplay of our stellar feedback model, the coupling of CR feedback, and the physics of gas cooling altogether influence the corresponding gas properties and are not cleanly separable i.e., these are the predictions of these CR transport models \textit{given} the FIRE-2 feedback and cooling physics and numerics. Changing the feedback and cooling prescriptions might lead to different results for the effect of the CR transport models on the synchrotron emission properties of simulated galaxies. The exact timing and prominence of these "blowout" events may also potentially depend on the gas resolution, which we will increase in future studies to $\sim$ 7000 M$_{\rm \odot}$, though we have checked the same CR transport variants for an intermediate-mass simulated galaxy (\texttt{m11f} in \citet{hopkins_testing_2021}, factor of $\sim$2 lower in halo mass than the simulations presented here) at a higher Lagrangian mass resolution of 12000 M$_{\rm \odot}$ and found similar results. The dynamical interaction of CRs again highlights the need for explicit evolution of CRs in galaxy formation simulations, as tracer particle or post-processing approaches to CR transport, for instance, popular methods like those of \texttt{GALPROP} \citep{strong1998} would by construction fail to capture these important effects.

Future work will include the exploration of more FIRE-3 simulations which vary CR transport and explicitly evolve CR(e) spectra beyond the "single-bin" simulations explored in this work. These FIRE-3 simulations will allow for the generation of more robust synchrotron predictions (i.e., spectral variation) that may generate new predictions for conducting observational tests of CR transport models. In a similar vein, multi-wavelength analysis of varied CR transport models, for example with spatial cross-correlations, may prove fruitful in generating more predictive constraints that can be tested against observations.

\section*{Acknowledgements}
We wish to recognize and acknowledge the past and present Gabrielino-Tongva people and their Indigenous lands upon which this research was conducted. Additionally, we thank the staff at our institutes, without whose endless efforts this work would not be possible during the ongoing pandemic. Support for SP and PFH was provided by NSF Research Grants 1911233, 20009234, 2108318, NSF CAREER grant 1455342, NASA grants 80NSSC18K0562, HST-AR-15800. GVP acknowledges support by NASA through the NASA Hubble Fellowship grant  \#HST-HF2-51444.001-A awarded  by  the  Space Telescope Science  Institute,  which  is  operated  by  the Association of Universities for Research in Astronomy, Incorporated, under NASA contract NAS5-26555. CBH is supported by NSF grant AAG-1911233 and NASA grants HST-AR-15800, HST-AR-16633, and HST-GO-16703.  Numerical calculations were run on the Caltech compute cluster "Wheeler," allocation AST21010 supported by the NSF and TACC, and NASA HEC SMD-16-7592. The Flatiron Institute is supported by the Simons Foundation. CAFG was supported by NSF through grants AST-2108230  and CAREER award AST-1652522; by NASA through grants 17-ATP17-0067 and 21-ATP21-0036; by STScI through grant HST-GO-16730.016-A; by CXO through grant TM2-23005X; and by the Research Corporation for Science Advancement through a Cottrell Scholar Award. ISB was supported by the DuBridge Postdoctoral Fellowship at Caltech. DK was supported by NSF grant AST2108314. KS acknowledges support from the Black Hole Initiative at Harvard University, which is funded by grants from the John Templeton Foundation and the Gordon and Betty Moore Foundation. This work was supported by NSF grant AST-2109127.
\section*{Data Availability}

The data supporting the plots within this article are available on reasonable request to the corresponding author. A public version of the GIZMO code is available at \url{http://www.tapir.caltech.edu/~phopkins/Site/GIZMO.html}. FIRE-2 simulations are publicly available \citep{Wetzel2022} at \url{http://flathub.flatironinstitute.org/fire}, though simulations including the physics of MHD and cosmic rays like those analyzed in this study are not yet publicly available. Additional data, including initial conditions and derived data products, are available at \url{https://fire.northwestern.edu/data/}.



\bibliographystyle{mnras}
\bibliography{Synch} 




\appendix


\bsp	
\label{lastpage}
\end{document}